\documentclass[12pt]{article}
\pdfoutput=1
\usepackage{euscript,amsmath,amsthm,amsfonts,amssymb}
\usepackage{mathtools}
\usepackage{subfig}
\usepackage{float}
\usepackage[margin=1in]{geometry}
\usepackage{graphicx}
\usepackage{outlines}
\usepackage[dvipsnames]{xcolor}
\usepackage{jheppub}
\usepackage{cleveref}

\usepackage{amssymb}
\usepackage{amsmath}
\usepackage{todonotes}
\usepackage{listings}

\usepackage[utf8]{inputenc}
\usepackage{color}
\usepackage{graphicx}
\usepackage{caption}
\usepackage{empheq}

\newcommand{\bra}[1]{\left\langle#1\right|}
\newcommand{\ket}[1]{\left|#1\right\rangle}

\newcommand{\su}[1]{\mathrm{SU}(#1)}
\newcommand{\sun}{\su{N}_1}

\newcommand{\nmod}{5 \mod{4}}
\newcommand{\mvec}[1]{\mathbf{#1}}

\numberwithin{equation}{section}

\title{Hypergraph States in $\su{N}_1$, $N$ odd prime, Chern-Simons Theory}
\author{Howard J.~Schnitzer}
\affiliation{Martin Fisher School of Physics, Brandeis University, Waltham MA 02453}
\preprint{BRX-TH-6672}
\emailAdd{schnitzr@brandeis.edu}

\abstract{Entropy cones for $SU(N)_{1}$ Chern-Simons theory are discussed. It is shown that stabilizer states can be constructed from topological operators in $SU(N)_{1}$ for $N$ odd prime, but not for $SU(N)_{K}$; $K \geq 2$. This implies that the topological entropy cone is properly contained in the stabilizer entropy cone for $SU(N)_{K}$; $K \geq 2$.

}

\abstract{Graph states and hypergraph states can be constructed from products of basic operations that appear in $\su{N}_1$. The level-rank dual of a theorem of Salton, Swingle, and Walter implies that these operations can be prepared topologically in the $n$-torus Hilbert space of Chern-Simons theory for $N \neq 5 \mod{4}$. 
    
For $\su{N}_1$, $N = 5 \mod{4}$, only stabilizer states can be prepared on the $n$-torus Hilbert space, which restricts the construction to graph states.}

\begin{document}
\maketitle
\flushbottom

\section{Introduction}
\label{sec:sec1}

The generalized Pauli group and Clifford operators \cite{schnitzer2019clifford,schnitzer2020n,schnitzer2020clifford} are obtained from products of basic operations of $\sun$ Chern-Simons theory for $N$ odd prime. Graph and hypergraph states \cite{steinhoff2017qudit,englbrecht2020symmetries,liu2020many,rossi2013quantum,hein2006entanglement} are described as products of such operations, which allow for explicit representations of graph and hypergraph states. For $N$ odd prime, $N \neq \nmod$, these operations can be obtained topologically in the $n$-torus Hilbert space of Chern-Simons theory as a result of the level rank dual \cite{mlawer1991group} of a theorem of Salton, Swingle, and Walter \cite{salton2017entanglement}. Thus graph and hypergraph states of $\sun$, $N \neq \nmod$, are topological on the $n$-torus Hilbert space of Chern-Simons theory.

For $\sun$, $N = \nmod$, only graph states are topological, as in this case only stabilizer states are obtained from the $n$-torus Hilbert space. Hypergraph states for $\sun$, $N = \nmod$, are therefore not topological in the above sense.

In Section \ref{sec:sec2} the construction of the generalized Pauli group and Clifford operations for $\sun$ is reviewed. Section \ref{sec:sec3} presents the construction of graph and hypergraph states in terms of basic operations of $\sun$. Section \ref{sec:sec4} is a discussion of related issues.

\section{$\su{d}_1$, $d$ odd prime}
\label{sec:sec2}

We first review the generalized Pauli group and Clifford operations for $\su{d}_1$, $d$ odd prime, following \cite{schnitzer2019clifford,schnitzer2020n,schnitzer2020clifford}.

\subsection{The $\su{d}_1$ Pauli group}
\label{sec:sec2a}

Representations of $\su{d}_1$ describing qudits are given by a single column Young tableau, with zero, one, \ldots, $(d-1)$ boxes. The fusion tensor of the theory is
\begin{equation}
    \label{eq:fusion}
    N_{ab}{}^c; \quad \quad a+b = c \mod{d}
\end{equation}
so that
\begin{equation}
    \label{eq:fusion2}
    N\ket{a}\ket{b} = \ket{a} \ket{a+b, \mod{d}}.
\end{equation}
The modular transformation matrix $S_{ab}$ satisfies
\begin{equation}
    \ket{a} = \sum_{b = 0}^{d-1} S_{ab} \ket{b}, \quad \quad a = 0 \ldots d-1.
\end{equation}
For $\omega$ a primitive $d$th root of unity
\begin{equation}
    \omega = \exp \left( \frac{2 \pi i}{d} \right)
\end{equation}
so that
\begin{equation}
    \label{eq:qft}
    S^* = \frac{1}{\sqrt{d}} \sum_{a = 0}^{d-1} \sum_{b = 0}^{d-1} \omega^{ab} \ket{a} \bra{b}
\end{equation}
which is the $d$-dimensional generalization of the Hadamard gate.

The Pauli operator $Z$ is given by
\begin{equation}
    Z_{ac} = \sum_{b = 0}^{d-1} S_{ab} \left( S^{\dagger}_{b+1,a} \right) \delta_{ac}
\end{equation}
so that
\begin{equation}
    Z = \sum_{a = 0}^{d-1} \omega^a \ket{a} \bra{a}.
\end{equation}
The Pauli operator $X$ is obtained from the fusion matrix, since
\begin{equation}
    \label{eq:paulix}
    N_{a,1}{}^b\ket{a} = \ket{a+1, \mod{d}}
\end{equation}
which is identical to
\begin{equation}
    X \ket{a} = \ket{a+1, \mod{d}}
\end{equation}
The single qudit Pauli group is the collection of operators
\begin{equation}
    \omega^r X^a Z^b; \quad \quad a,b,r \in \mathbb{Z}_d.
\end{equation}
Thus the one-qudit Pauli group is constructed from basic operations of $\su{d}_1$, $d$ odd.

The $n$-qudit Pauli group is obtained from products of operators of the one-qudit Pauli group. That is
\begin{equation}
    X^{\mvec{a}} Z^{\mvec{b}} = X^{a_1} Z^{b_1} \otimes X^{a_2} Z^{b_2} \otimes \cdots \otimes X^{a_n} Z^{b_n}
\end{equation}
The operator $X^{\mvec{a}} Z^{\mvec{b}}$, along with all scalar multiples thereof,
\begin{equation}
    \left\{ \omega^c X^{\mvec{a}} Z^{\mvec{b}} \ | \ c \in \mathbb{Z}_d \right\}
\end{equation}
defines the $n$-qudit Pauli group.

\subsection{$\su{d}_1$ Clifford operators, $d$ odd}

The necessary gates for the single-qudit Clifford operators are \cite{schnitzer2019clifford,schnitzer2020n,schnitzer2020clifford} i) the QFT gate, Eq.~\eqref{eq:qft}, and ii) the phase gate
\begin{equation}
    P \ket{j} = \omega^{j(j-1)/2} \ket{j}.
\end{equation}
The multi-qudit Clifford operators are obtained from the generalizations of \eqref{eq:qft} and \eqref{eq:paulix} as well as the SUM gate,
\begin{align}
    C_{\mathrm{SUM}} \ket{i} \ket{j} &= N \ket{i} \ket{j} \nonumber \\
    &= \ket{i} \ket{i+j, \mod{d}}
\end{align}

\section{Graph and hypergraph states}
\label{sec:sec3}

\subsection{Graph states}
\label{sec:sec3a}

There are many equivalent constructions of graph states \cite{steinhoff2017qudit,englbrecht2020symmetries,liu2020many,rossi2013quantum,hein2006entanglement}. We follow arxiv:1612.06418 for a definition of qudit graph states. The multigraph is $G = (V,E)$, with vertices $V$ and edges $E$, where an edge has multiplicity $m_e \in \mathbb{Z}_d$. To $G$ associate a state $\ket{G}$ such that to each vertex $i \in V$, there is a local state
\begin{equation}
    \ket{+} = \ket{p_0} = \frac{1}{\sqrt{d}} \sum_{q = 0}^{d-1} \ket{q}
\end{equation}
Recall that the Hadamard gate generalizes to \eqref{eq:qft}, so that
\begin{align}
    S^* \ket{0} &= \frac{1}{\sqrt{d}} \sum_{q = 0}^{d-1} \ket{q} \nonumber \\
    &= \ket{+} = \ket{p_0}.
\end{align}

To each edge $e = \{i,j\}$ apply the unitary
\begin{equation}
    Z_e^{m_e} = \sum_{q_i = 0}^{d-1} \ket{q_i} \bra{q_i} \otimes \left(Z_j^{m_e}\right)^{q_i}
\end{equation}
to the state
\begin{equation}
    \ket{+}^V = \bigotimes_{i \in V} \ket{+}_i
\end{equation}

The graph state is
\begin{align}
    \ket{G} &= \prod_{e \in E} Z_e^{m_e} \ket{+}^V \\
        &= \prod_{e \in E} Z_e^{m_e} \bigotimes_{i \in V} \ket{+}_i
\end{align}
The level-rank dual \cite{mlawer1991group} of Theorem 1 of Salton, Swingle, and Walter \cite{salton2017entanglement} for $d$ odd prime implies that the graph state $\ket{G}$ can be constructed from topological operations on the $n$-torus Hilbert space of Chern-Simons $\su{d}_1$ by means of the operations detailed in Section \ref{sec:sec2}. Every stabilizer state is LC equivalent to a graph state, while the Clifford group enables conversion between different multigraphs \cite{steinhoff2017qudit,englbrecht2020symmetries,liu2020many,rossi2013quantum,hein2006entanglement}.

\subsection{Hypergraph states}

We again follow arxiv:1612.06418 for the construction of qudit multi-hypergraph states. Given a multi-hypergraph $H = (V,E)$, associate a quantum state $\ket{H}$, with $m_e \in \mathbb{Z}_d$ the multiplicity of the hyperedge $e$. To each vertex $i \in V$, associate a local state
\begin{align}
    \ket{+} &= \frac{1}{\sqrt{d}} \sum_{q=0}^{d-1} \ket{q} \nonumber \\
        &= S^* \ket{0}
\end{align}
To each hyperedge $e \in E$, with multiplicity $m_e$, apply the controlled unitary $Z_e^{m_e}$ to the state
\begin{equation}
    \ket{+}^V = \bigotimes_{i \in V} \ket{+}_i
\end{equation}
The hypergraph state is
\begin{align}
    \ket{H} = \prod_{e \in E} Z_e^{m_e} \ket{+}^V
\end{align}
The elementary hypergraph state is
\begin{align}
    \ket{H} = \sum_{q=0}^{d-1} \ket{q} \bra{q} \otimes \left( Z_{e \backslash \{1\}}^{m_e} \right)^q \ket{+}^V
\end{align}
For $d$ prime, all $n$-elementary hypergraph states are equivalent under SLOCC.

Hypergraph and graph states admit a representation in terms of Boolean functions,
\begin{equation}
    \ket{H} = \sum_{q = 0}^{d-1} \omega^{f(q)} \ket{q}
\end{equation}
with $f: \mathbb{Z}_d^n \rightarrow \mathbb{Z}_d$, where
\begin{equation}
    \label{eq:boolfunc}
    f(x) = \sum_{\substack{i_1,\ldots,i_k \in V \\ \{ i_1,\ldots,i_k \} \in E}} x_{i_1} \cdots x_{i_k}
\end{equation}
For graph states, $f(x)$ is quadratic, i.e.
\begin{equation}
    f(x) = \sum_{\substack{i_1,i_2 \in V \\ \{ i_1,i_2 \} \in E}} x_{i_1} x_{i_2}
\end{equation}
while for $f(x)$ cubic or higher, $\ket{H}$ is a hypergraph state. Therefore, for quadratic $f(x)$, one has a representation of stabilizer states, up to LC equivalence. For $f(x)$ cubic or higher, $\ket{H}$ represents hypergraph states which contain ``magic'' states. Examples of magic states are the $\mathrm{CCZ}$ state and Toffoli states, constructed from appropriate gates \cite{heyfron2019quantum,paetznick2013universal,o2017quantum,schnitzer2020clifford,eastin2013distilling,jones2013low,howard2012qudit,biswal2019techniques,haah2018codes}. Thus
\begin{equation}
    \mathrm{CCZ} \ket{x_1 x_2 x_3} = \omega^{x_1 x_2 x_3} \ket{x_1 x_2 x_3}
\end{equation}
with
\begin{equation}
    \ket{\mathrm{CCZ}} = \mathrm{CCZ} \ket{+ \otimes^3}
\end{equation}
as an example of a magic hypergraph state. Similarly
\begin{equation}
    \ket{\mathrm{Toff}} = \mathrm{Toff} \ket{+ \otimes^3}
\end{equation}
where the Toffoli gate can be expressed in terms of the fusion matrix \eqref{eq:fusion2}. Explicitly,
\begin{align}
    \mathrm{Toff} \ket{i,j,k} &= N_{ij,k}{}^{(ij+k)} \nonumber \\
        &= \ket{i,j,ij+k, \mod{d}}
\end{align}
where the Young tableau for $(ij)$ has $i+j$ vertical boxes, $\mod{d}$.

For $\su{d}_1$, $d$ odd prime, $d \neq \nmod$, both graph and hypergraph states can be obtained from operators which can be constructed from products of topological operations on the $n$-torus Hilbert space \cite{salton2017entanglement}.

For $\su{d}_1$, $d = \nmod$, the topological argument does not apply, since in this case Theorem 1 of Salton \emph{et al} \cite{salton2017entanglement} implies that \emph{only} stabilizer states can be constructed on the $n$-torus Hilbert space. Thus, graph states can be so constructed but not hypergraph states with cubic or higher functions \eqref{eq:boolfunc}.

\section{Discussion}
\label{sec:sec4}

It was shown above that graph states and hypergraph states for $\su{d}_1$, $d$ odd, can be constructed from the basic operations $N_{ab}{}^c$, $S_{ab}$, $Z$, and $X$ of $\su{d}_1$ Chern-Simons theory. For $d$ odd prime, $d \neq \nmod$, these operations can be constructed topologically on the $n$-torus Hilbert space of Chern-Simons theory. A subset of hypergraph states are ``magic.'' For $d \neq \nmod$, they are topological in the above sense.

Fliss \cite{fliss2020knots} has studied knot and link states of $\su{2}_d$ Chern-Simons theory, and has shown that knot and link states are generically magical. However for $\mathrm{U}(1)_d$, magic is absent for all knot and link states. Since $\mathrm{U}(1)_d$ is level-rank dual to $\su{d}_1$, the knot and link states for this theory also have zero magic \cite{fliss2020knots,balasubramanian2017multi,balasubramanian2018entanglement},\cite{Jain_2020}.

There is a great deal of recent interest in magic states \cite{white2020conformal,white2020mana,liu2020many,veitch2014resource}. One feature that deserves further study is to understand which magic states are topological. For example, universal topological computing is possible for $\su{2}_3$ \cite{freedman2002modular} and $\su{3}_2$ \cite{schnitzer2018level} Chern-Simons theory. Implicitly this implies that magic states are present in these theories, presumably due to the braiding operations. It would be interesting to make this explicit.

\acknowledgments
\addcontentsline{toc}{section}{Acknowledgments}

We thank Zi-Wen Liu for clarifying the difference between hypergraph / stabilizer cones \cite{bao2020quantum,walter2020hypergraph,bao2020gap,bao2020bit}, and hypergraph / stabilizer states, and for emphasizing that universal quantum computation models for $\su{2}_3$ and $\su{3}_2$ have gates / actions which are magical.

We are grateful to Greg Bentsen and Isaac Cohen for their assistance in the preparation of the paper.

\addcontentsline{toc}{section}{References}

\nocite{*}
\bibliographystyle{JHEP}
\bibliography{main}

\end{document}